\documentclass{aa}
\usepackage{graphicx}
\usepackage{psfig}
\begin{document}


\title{Cross-correlation between the soft X-ray background and SZ Sky}

\author{Ling-Mei Cheng\inst{1,2} \and Xiang-Ping Wu\inst{1}
        \and Asantha Cooray\inst{3}}

\offprints{L.-M. Cheng}
\mail{clm@class1.bao.ac.cn}
\institute{National Astronomical Observatories,
       Chinese Academy of Sciences, Beijing 100012, China; 
\and 
  Department of Astronomy, Beijing Normal University,  
  Beijing 100875, China; 
\and 
California Institute of Technology, 
  Mail Code 130-33, Pasadena, CA 91125} 

\date{Received 00 March, 2003; accepted}

   \titlerunning{SXRB and SZ Cross-correlation}
   \authorrunning{Cheng et al.}

\abstract{
While both X-ray emission and Sunyaev-Zel'dovich (SZ) temperature 
fluctuations are generated by the warm-hot gas in dark matter halos, the 
two observables have different dependence on the underlying physical 
properties, including the gas distribution.
A cross-correlation between the soft X-ray background (SXRB) and
the SZ sky may allow an additional probe on the distribution 
of warm-hot gas at intermediate angular scales and redshifts complementing 
studies involving clustering within SXRB and SZ separately. Using a halo 
approach, we investigate this cross-correlation analytically.
The two contributions are correlated mildly with a correlation 
coefficient of $\sim0.3$, and this relatively low correlation presents 
a significant challenge for its detection. The correlation, at small 
angular scales, is affected by the presence of radiative cooling or 
preheating and provides a probe on the thermal history of the hot 
gas in dark halos. While the correlation remains 
undetectable with CMB data from the WMAP satellite and 
X-ray background data from existing catalogs, upcoming observations 
with CMB missions such as Planck, for the SZ side, and an improved 
X-ray map of the large scale structure, such as the one planned with
DUET mission, may provide a first opportunity for a reliable detection 
of this cross-correlation. 

}
\maketitle
\section{Introduction}

The best current census of baryons in the universe conducted at  
high and low redshifts reveals that a considerably large fraction
of baryons in the local universe is still missing (Fukugita et al.
1998). Hydrodynamical simulations of structure formation 
suggest that the missing baryons may exist in the form of warm-hot
intergalactic medium (WHIM) with temperatures of $T\sim 10^5-10^7$K 
(Cen \& Ostriker 1999; Dav\'e et al. 2001).  This arises because
baryons can be gravitationally heated and adiabatically compressed
when they fall into large-scale structures, 
including collapsed dark matter halos.
However, while there has been observational evidence for presence of missing 
baryons associated with large-scale structures at low redshifts, which
includes the degree-scale X-ray filaments 
(Scharf et al. 2000; Zappacosta et al. 2002), the soft X-ray excess 
emission in the vicinity of nearby clusters (Nevalainen et al.
2003; Kaastra et al. 2003), the resonant absorption
lines of local warm gas towards distant AGNs/QSOs
(e.g. Fang et al. 2002; Nicastro et al. 2002, 2003),
the statistical confidences related to these detections remain poor and 
majority of the baryons still escape the direct detection.

In addition to X-ray emission, missing baryons manifest themselves 
through the inverse-Compton scattering of
cosmic microwave background (CMB) photons. The latter is known as 
the Sunyaev-Zel'dovich (SZ) effect. Indeed,  massive groups and clusters 
that serve as a reservoir of the WHIM are very luminous sources 
in both X-ray and SZ maps. The diffuse WHIM in poor groups and 
filamentary structures associated with the large-scale ``cosmic web'', 
may not be strong enough to allow direct detection in individual cases. 
This gas, however, may make a significant contribution to the SXRB and 
CMB temperature fluctuation background related to the SZ effect, 
provided that the known sources such as AGNs and nearby, rich clusters can 
be removed from the SXRB and SZ sky. 
Recall that a considerably fraction ($80$--$90\%$) of 
the SXRB has been resolved into discrete sources (see Xue \& Wu 2003 for
a recent summary). 
To extract the presence of missing baryons and reconstruct the content 
and the distribution,  one needs to rely on certain statistical approaches. 
These include the two-point auto-correlation function of the SXRB or SZ sky, 
or the cross-correlation of SXRB and/or SZ map with galaxies, groups 
and clusters (e.g. Soltan et al. 2001, 2002; Zhang \& Pen 2003; 
Wu \& Xue 2003). Another possibility discussed here 
is the cross-correlation between maps of SXRB and SZ effect.

As is known, X-ray measurement is a sensitive probe of the hot gas 
that is located in and near central regions of dark matter halos, 
mainly groups and clusters, and in the local universe. 
Recall that the X-ray emissivity within the framework of bremmstrahlung 
is proportional to the  square of the electron density, and 
the X-ray flux is inversely proportional to the square of the distance from us.
This compares to the SZ effect, which reveals a more extended 
distribution of hot gas in dark halos as a result of the linear
dependence of the electron density and out to high 
redshifts because of the rather weak dependence on distance.
Therefore, it is expected that the cross-correlation between 
the SXRB from diffuse gas and the SZ sky may allow one to explore the 
distribution and evolution of the gas at intermediate scales and redshifts.

To estimate the extent to which this cross-correlation is present 
and whether it can be detected, we make use of a halo approach
(e.g., Cooray \& Sheth 2002) with gas assumed to either trace 
dark matter or follow the $\beta$-profile. 
We then investigate how non-gravitational heating and radiative cooling can
modify the cross power spectrum following discussions related to these
effects in Wu \& Xue (2003).  We also discuss another interesting application 
of the SXRB and SZ cross-correlation. Since the SXRB traces the square of the 
number density of electrons,
while the SZ contribution is only sensitive to the integrated number density, 
any clumping of the gas distribution, such that the mean of the squared gas 
distribution is higher than the square of the mean, the cross-correlation 
between SXRB and SZ will be augmented at  angular scales corresponding to the
clumping of gas. Thus, any reliable detection of the SXRB and SZ 
cross-correlation can then be used to understand the clumped nature of the 
gas distribution, which is an important aspect of the WHIM and
cannot be easily obtained by other means.

To illustrate our results, we adopt a flat cosmological model 
($\Lambda$CDM) with the best fit parameters determined by WMAP
(Spergel et al. 2003): $\Omega_M=0.27$, $\Omega_{\Lambda}=0.73$, 
$\Omega_b h^2=0.0224$, $h=0.71$, $\sigma_8=0.84$ and $n_s=0.93$. 
During the preparation of this work, a paper by
Diego et al. (2003) appeared claiming no detection of the 
cross-correlation between CMB data, 
as obtained by the WMAP satellite, and X-ray background data obtained 
by a map of the ROSAT soft X-ray emission.
This lack of CMB-SXRB cross-correlation may be attributed to 
either a smaller value of $\sigma_8$ or the relatively weak signals
at large angular scales. While the initial claim 
for a detection has disappeared, we suggest that the
upcoming Planck mission will allow a first detection of the SZ-SXRB 
cross-correlation. The advantage over current data is that
with Planck, one can use multifrequency information to extract a 
separate map of the large scale structure SZ effect
(Cooray et al. 2000), which can then be cross-correlated directly
with an X-ray map. The current approach, involving
WMAP data, is not likely to be useful given that the fluctuations 
related to SZ is dominated by the primordial fluctuations
of CMB related to physics at last scattering, such as the acoustic 
peak structure.

\section{SXRB and SZ cross-correlation}

\subsection{SXRB and SZ effect}

The thermal SZ effect along the direction ${\mbox{\boldmath $\theta$}}$ 
due to the hot gas inside a halo can be evaluated following
\begin{eqnarray}
\frac{\Delta T({\mbox{\boldmath $\theta$}})}
        {T_{\rm CMB}}&=& g_{\nu}(x)y({\mbox{\boldmath $\theta$}});\\
y({\mbox{\boldmath $\theta$}})&=&
       \int n_e \sigma_T \left(\frac{k_BT}{m_e c^2}\right)\rm d\chi;\\
g_{\nu}(x)&=&\frac{x^2 e^x}{(e^x-1)^2}\left(4-x \coth\frac{x}{2}\right),
\end{eqnarray}
where $x=h_p\nu/kT_{\rm CMB}$ is the dimensionless frequency, 
$T_{\rm CMB}$ is the temperature of the present CMB, 
and the integral is performed along the line of sight, $\chi$.

In terms of bremmstrahlung emission, the X-ray surface brightness 
distribution in direction ${\mbox{\boldmath $\theta$}}$ is given by
\begin{equation}
S_{\rm X}({\mbox{\boldmath $\theta$}})=
        \frac{1}{4\pi(1+z)^4} \int n_e n_H \Lambda(T,Z) d\chi,
\end{equation}
where $n_H$ is the number density of hydrogen, and $\Lambda(T,Z)$ is the 
cooling function in a given energy band that is calculated using the 
Raymond-Smith(1977) code with a metallicity of $Z=0.3Z_\odot (t/t_0)$,
and $t_0$ is the present age of the universe.
The mean SXRB brightness $S_X$ is calculated through 
\begin{equation}
\langle S_{\rm X}\rangle=
               \int dz \frac{dV}{d \Omega dz}\int dM 
              \frac{d^2N}{dM dV} \left[\frac{L_X(M,z)}{4\pi D_L^2(z)}\right],
\end{equation}
where $L_{\rm X}(M,z)$ is the total X-ray luminosity of a halo of mass $M$ at 
redshift $z$, $D_L$ is the luminosity distance, and
$d^2N/dMdV$ is the comoving number density of dark halos. For the latter
we adopt the mass function of Jenkins et al. (2001).

We follow two approaches to describe the electron distribution within halos.
First, gas is assumed to trace dark matter such that
\begin{equation}
n_e=\frac{f_b}{\mu_e m_p}\rho_{\rm DM},
\end{equation}
in which we have introduced the universal baryon fraction 
$f_b=\Omega_b/\Omega_M$, and $\mu_e=1.13$ is the mean electron weight.
We adopt the universal density profile as suggested by 
cosmological numerical simulations (Navarro et al. 1995; NFW)
to describe the dark matter distribution, $\rho_{\rm DM}$, in halos
\begin{equation}
\rho_{\rm DM}(r)=\frac{\delta_{\rm ch}\rho_{\rm crit}}
          {(r/r_{\rm s})(1+r/r_{\rm s})^2},
\end{equation}
where $\delta_{\rm ch}$ and $r_{\rm s}$ are the characteristic 
density and length of the halo, respectively, 
which can be fixed through the so-called concentration 
parameter $c=r_{\rm vir}/r_{\rm s}$ using the empirical 
fitting formula found by numerical simulations (Bullock et al. 2001)
\begin{equation}
c=\frac{10}{1+z}\left(\frac{M}{2.1\times10^{13}M_{\odot}}\right)^{-0.14}.
\end{equation}
The second approach involves the $\beta$-model with
\begin{equation}
n_e=\frac{n_{\rm e0}}{\left[1+(r^2/r_c^2)\right]^{3\beta/2}}.
\end{equation}
We fix the $\beta$ value to $\beta=2/3$ and specify  
the core radius by $r_c=0.1r_{\rm vir}$. The normalization $n_{\rm e0}$
is determined using the universal baryon fraction. 
Finally, we need to specify the temperature profile $T(r)$.
For the first model, $T(r)$ can be obtained in principle by solving the 
equation of hydrostatic equilibrium, while for the second one   
either the gravitational potential of dark matter or the equation of state 
for the gas must be given in order to work out $T(r)$. 
For simplicity, we now assume an isothermal model for the gas 
distribution in both approaches 
and specify its temperature in terms of virial theorem 
(e.g. Bryan and Norman 1998):
\begin{equation}
kT=1.39\,{\rm keV} f_T 
   \left(\frac{M}{10^{15}M_{\odot}}\right)^{2/3}(h^2E^2\Delta_c)^{1/3}
\end{equation}
where $\Delta_c$ denotes the overdensity parameter, and 
$f_T$ is the normalization factor which will be fixed 
to be $f_T=0.8$ in the following 
evaluation. While the isothermality may certainly 
introduce some uncertainties in our numerical predictions, this
assumption should be good enough to provide an approximate estimate of 
to what extent the SXRB and SZ sky may correlate with each other 
at small scales. As an example, the typical central electron densities 
in the $\beta$-model turn to be $6.6\times10^{-3}$ cm$^{-3}$ and 
$0.015$ cm$^{-3}$ for a group of $M=5\times10^{13}$ $M_{\odot}$ and
a rich cluster of $M=5\times10^{15}$ $M_{\odot}$, respectively, at
present epoch.

\subsection{Power spectra}

Following halo approach to large scale structure clustering,
the angular cross power spectrum of the SXRB-SZ correlation
can be separated into
the Poisson term $C_{\ell}^P$ and the clustering term $C_{\ell}^C$:
\begin{eqnarray}
  \label{eq:Cp}
  C_{\ell}^{P} &=& g_{\nu}(x)\int_0^{z_{\rm dec}} dz \frac{dV}{dzd\Omega}
             \int_{M_{\rm min}}^{\infty} dM
                \frac{d^2N(M,z)}{dMdV} \nonumber \\
&&\quad \quad \quad \times |y_{\ell}(M,z)s_{\ell}(M,z)|,
\end{eqnarray}
and
\begin{eqnarray}
  \label{eq:Cc}
  C_l^{C} &=& g_{\nu}(x)\int_0^{z_{\rm dec}} dz \frac{dV}{dzd\Omega} 
                P(k=\ell/D,z) \nonumber \\
	    & &	 \times
                \left [\int_{M_{\rm min}}^{\infty} dM
                \frac{d^2N(M,z)}{dMdV} b(M,z) y_{\ell}(M,z) \right] 
                                              \nonumber \\
		&& \times
	        \left [\int_{M_{\rm min}}^{\infty} dM
                \frac{d^2N(M,z)}{dMdV} b(M,z) s_{\ell}(M,z) \right],
\end{eqnarray}
where $z_{\rm dec}\approx1000$ is the CMB photon decoupling redshift, 
$D$ is the comoving angular diameter distance to the halo of mass $M$ 
at $z$, and $y_{\ell}$ and $s_{\ell}$ are the Fourier transforms of 
the Compton $y$-parameter and the SXRB fluctuation, 
$s(\theta)\equiv[S_{X}(\theta)-\langle S_{X}\rangle]/\langle S_{X}\rangle$, 
respectively.
Here, $b(M,z)$ is the bias parameter, for which we use
the analytic prescription of Mo \& White (1996). We take
the minimum halo mass to be $M_{\rm min}=10^{12}M_\odot$. 
Our final result is unaffected by this choice since both the SZ sky
and SXRB are dominated by massive halos, and the shallower
gravitational potential of galactic halos can hardly preserve
the hot gas because of preheating or feedback process of 
star formation.
Replacing $g_{\nu}(x)y_{\ell}$ (or $s_{\ell}$) by $s_{\ell}$ 
[or $g_{\nu}(x)y_{\ell}$] in equations (11) and (12), we can get 
power spectra of the SXRB (or SZ map).
Finally, the cross-correlation coefficient is calculated by
\begin{equation}
r_{\ell} = \frac{C_{\ell}^{\rm SXRB-SZ}}
           {\sqrt{C_{\ell}^{\rm SZ} C_{\ell}^{\rm SXRB}}} \,,
\end{equation}
which quantitatively indicates the strength of the cross-correlation.

\subsection{Non-gravitational effect}

Presence of non-gravitational effect on the distribution and
global properties of hot gas in groups and clusters has been
firmly established over the past few years.  Here
we use a simple phenomenological approach to demonstrate how
the SXRB-SZ cross power spectrum is modified by the  
non-gravitational processes such as preheating or radiative cooling,
both of which tend to flatten the gas distribution in the central
regions of groups and clusters and become indistinguishable in 
the explanation of the observed X-ray properties of groups and
clusters. We refer the reader to 
the recent work of Xue \& Wu (2003) for a set of analytic models
of preheating and cooling processes and the predicted properties 
of the WHIM.  The basic procedures are summarized below:

(1) Preheating model. We begin with the gas distribution 
predicted by self-similar model and then raise the entropy distribution 
of the WHIM in dark halos, defined as $S=kT/n_e^{2/3}$, by a certain level 
$S_{\rm floor}$ which is fixed to $120$ keV cm$^2$ as 
suggested by X-ray measurements of groups and clusters
(Ponman et al. 1999; but see Ponman et al. 2003). 
The new distribution of the WHIM can be obtained by solving the 
equation of hydrostatic equilibrium.

(2) Cooling model. A certain amount of WHIM in the central regions 
of groups and clusters will be removed from the hot phase if its
cooling time is shorter than the cosmic age, and higher-entropy
gas at large radii would then flow inward.  
Following the prescription of Voit \& Bryan (2001)
and Wu \& Xue (2002), we first estimate the total cooled mass of WHIM 
within the cooling radius by combining energy conservation and setting 
the cooling time to equal the cosmic age. We then derive the new 
distribution of the WHIM after cooling by solving the equation 
of hydrostatic equilibrium under the conservation of total baryonic 
mass and entropy.

\subsection{Gas Clumping}

In order to make an estimate of the effect related to WHIM clumping spatially,
we define the clumping parameter as
\begin{equation}
C(r)=\frac{<n_e^2(r)>}{<n_e(r)>^2} \, ,
\end{equation} 
which describes the excess gas distribution related to the mean. Note in 
this definition we ignored the unhomogeneity of temperature. Since the
X-ray emission scales as square of the number density, any clumping affects 
the SXRB while SZ remains unaffected unless the density variation is 
completely compensated by the temperature modulation.
While one can potentially use the SXRB 
information alone to understand clumping, this requires prior knowledge on 
the mean of the gas distribution and this is readily available from the SZ 
side. Thus, a combined study of SZ and SXRB, such as the cross-correlation of 
the two, can be used to extract information related to any potential clumping 
of the gas distribution.

\section{Results}

In Figure 1, we show the power spectra of 
SXRB and SZ effect; the two are calculated 
in the energy band 0.5-2.0 keV and at the frequency $\nu=30$ GHz, 
respectively. Our result also holds for other observing frequencies
when the frequency dependence term $g(\nu)$ in the SZ effect 
is correspondingly corrected.
It is apparent that the SXRB shows a stronger power 
at $\ell>10^4$ than the SZ map, which arises simply from
the different dependence of the SXRB and SZ effect on the gas density:
The former varies as $n_e^2$ while the latter goes as $n_e$. 
Employment of a $\beta$ model with $\beta=2/3$ leads to a 
significant drop of the power spectra at large $\ell$,
especially for the SXRB, which can be attributed to 
the relatively flat core radius in the gas distribution 
when compared to the NFW-like profile.
\begin{center}
\begin{figure*}
   \resizebox{\hsize}{!}{\includegraphics{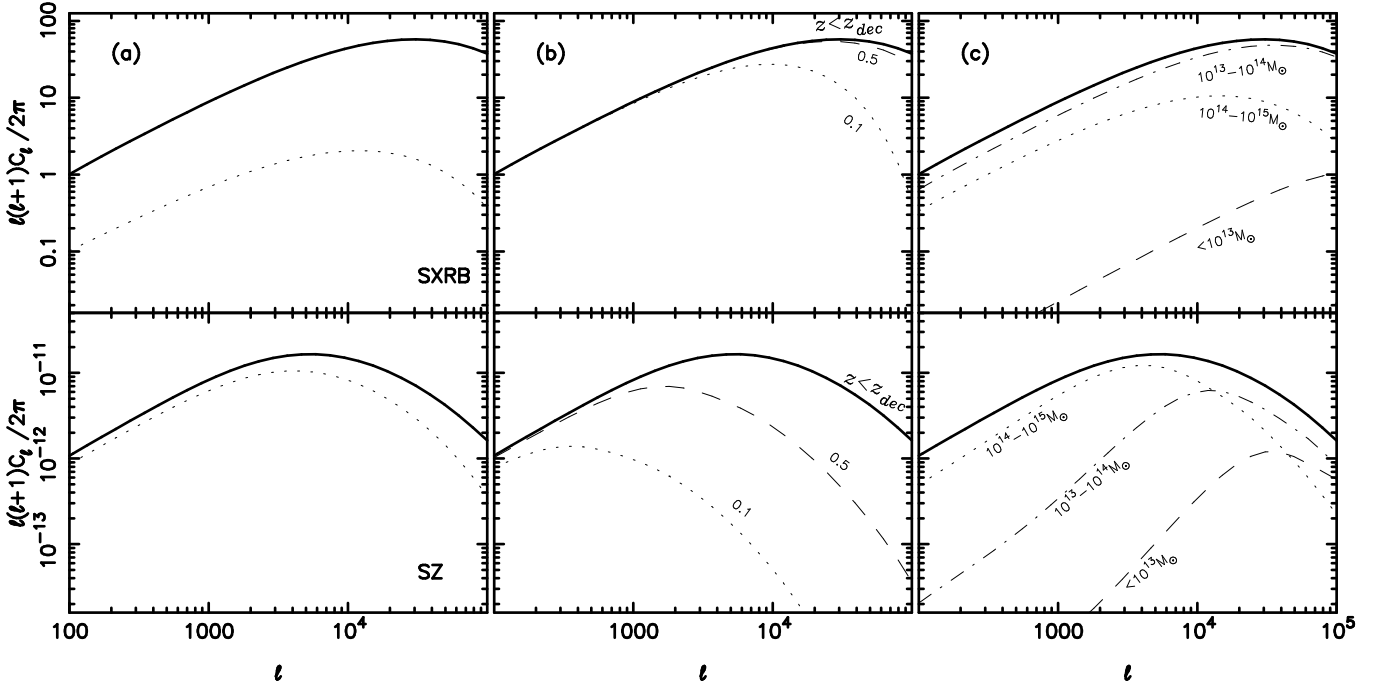}}
\caption{The angular power spectra of 
the SXRB (upper panel) and the SZ sky (lower panel) measured in
the energy band 0.5-2.0 keV and at frequency $\nu=30$ GHz, respectively.
In (a), two models are assumed for the gas distribution inside dark halos:
the self-similar model (NFW-like model; solid lines) and  
$\beta$ model with $\beta=2/3$ and 
$r_c=0.1r_{\rm vir}$ (dotted lines). In (b), we show 
contributions from halos at different redshifts while in (c), we show
contributions from halos separated in mass.}
\label{f1}
\end{figure*}
\end{center}

In Figure 1(b), 
we illustrate contributions of halos in different redshift ranges
to the SXRB and SZ power spectra. As it is expected, 
the SXRB power spectrum at $\ell<10^5$ is dominated by nearby halos within 
$z<0.5$, and high-redshift halos only make a minor contribution 
to the power spectrum at $\ell$ up to $10^5$, 
which is due to the well-known inverse dependence of X-ray flux 
on the square of cosmic distances to halos. On the contrary,
because the SZ effect reflects the thermal energy of hot gas intrinsic
to halos,  the SZ power spectrum, at small angular scales 
corresponding to $\ell>10^3$, is governed by halos at $z>0.5$.

In Figure 1(c), we show the dependence of SXRB and SZ power spectra 
on halo mass. It appears that massive clusters with 
$M>10^{14}$ $M_{\odot}$ determine the SZ 
power spectrum at $\ell<10^4$, and at sub arcminute scales when
$\ell\sim10^5$, poor clusters and groups 
with $M\sim10^{13}$--$10^{14}$ $M_{\odot}$ starts to contribute.
This compares to the SXRB power spectrum,
when $\ell>10^2$, that is entirely dominated by low mass groups. 
Inclusion of non-gravitational heating or radiative 
cooling processes may alter this mass-dependence (see Figure~3).

In Figure~2, we show the angular cross power spectrum between the 
SXRB and the SZ effect calculated with NFW and $\beta$ models. Note that
we have dropped the negative sign in the power spectrum arising from
$g_{\nu}(x)<0$ for $\nu=30$ GHz.   
The cross power spectrum in the case of NFW model peaks at 
$\ell\approx 10^4$, between peak locations of
SXRB and SZ power spectra.  The same 
conclusion also applies to the $\beta$ model, though the peak now appears
at a slightly smaller value of $\ell$. Moreover, 
at $\ell$ values beyond $10^2$,
the contribution related to the halo-halo clustering term is insignificant.
The contribution of halos at different redshifts and with different masses
to the SXRB-SZ cross power spectrum is shown in Figure 2(b) and 2(c). 
Essentially, at $\ell<10^4$, the power spectrum is governed by nearby 
clusters and groups with masses in the range
$M>10^{14}$ $M_{\odot}$ within $z=0.5$.
In the range of $10^4<\ell<10^5$,  groups with  masses in the range of
$10^{13}<M<10^{14}$ $M_{\odot}$ and at redshifts greater than 0.5 begin to 
dominate the power spectrum. Unlike the SXRB power spectrum,
the cross-correlation of SXRB and SZ contributions
at $\ell>10^4$ is sensitive to clusters and groups  
at high redshifts. This, of course, is a combined 
result of the different redshift dependence of the SXRB and 
SZ power spectra.

\begin{figure}
	\psfig{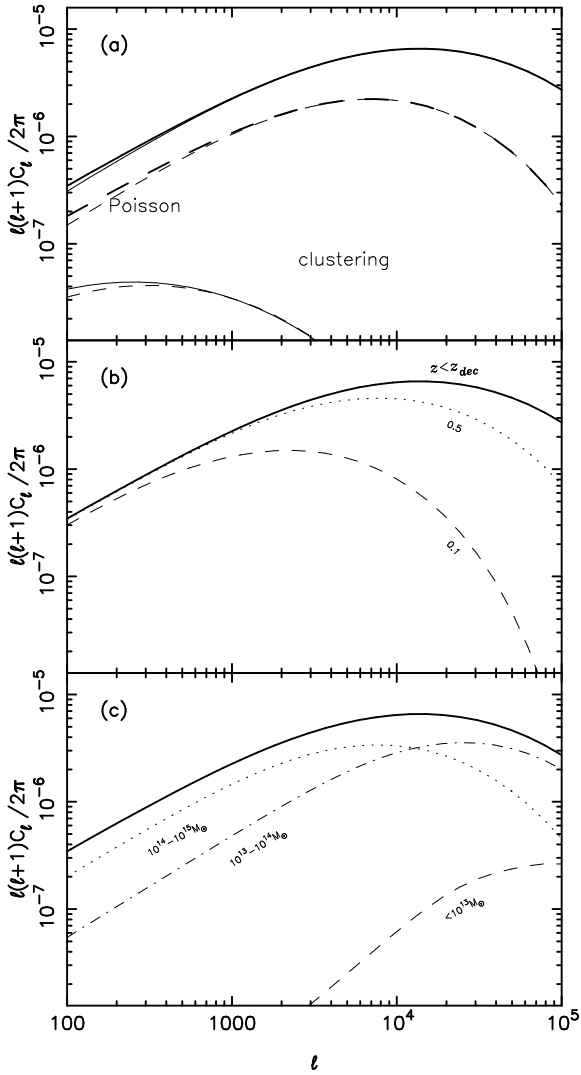}
\caption{(a): The expected power spectra of the SXRB-SZ cross-correlation for
the NFW-like profile (solid lines) and  $\beta$ model (dashed lines). 
Contributions from  the Poisson and clustering distributions of dark 
halos are explicitly shown (thin lines). In (b) and (c), 
contributions from halos at different redshifts and varying masses 
are shown, respectively.}
\label{f2}
\end{figure}

The effects of preheating and radiative cooling on the 
power spectra are illustrated in Figure 3, together with 
the auto-correlation power spectra of SXRB and SZ sky. 
As compared with the prediction of self-similar
models, inclusion of preheating or cooling leads to a significant 
modification of the shape of the cross power spectrum at small 
angular scales, which can be attributed to the flattened distribution of 
the hot gas in the central regions of dark halos produced by 
preheating and/or radiative cooling. 
Actually, the effect of nongravitational process is somewhat equivalent 
to the excision of a certain inner region in groups and clusters 
in the evaluation of SZ effect, which always leads to
a decrease of the SZ power-spectrum amplitude at small scales
(Komatsu \& Seljak 2002).
While our predicted power spectra of the SZ sky and SXRB 
with preheating or cooling are roughly consistent with 
the previous studies of da Silva et al. (2001), Zhang \& Wu (2003), 
Zhang \& Pen (2003), and Wu \& Xue (2003), the extent to which 
the SZ (or SXRB) power spectrum is modified by nongravitational effect 
is still uncertain. For example, using high resolution hydrodynamic
simulations White et al. (2002) found that 
nongravitational effect on the SZ power spectrum is only minor.
A combination of the future high-resolution data of SXRB-SXRB, 
SZ-SZ and/or SXRB-SZ correlations at different scales may allow
us to clarify the issue.

\begin{figure}
        \psfig{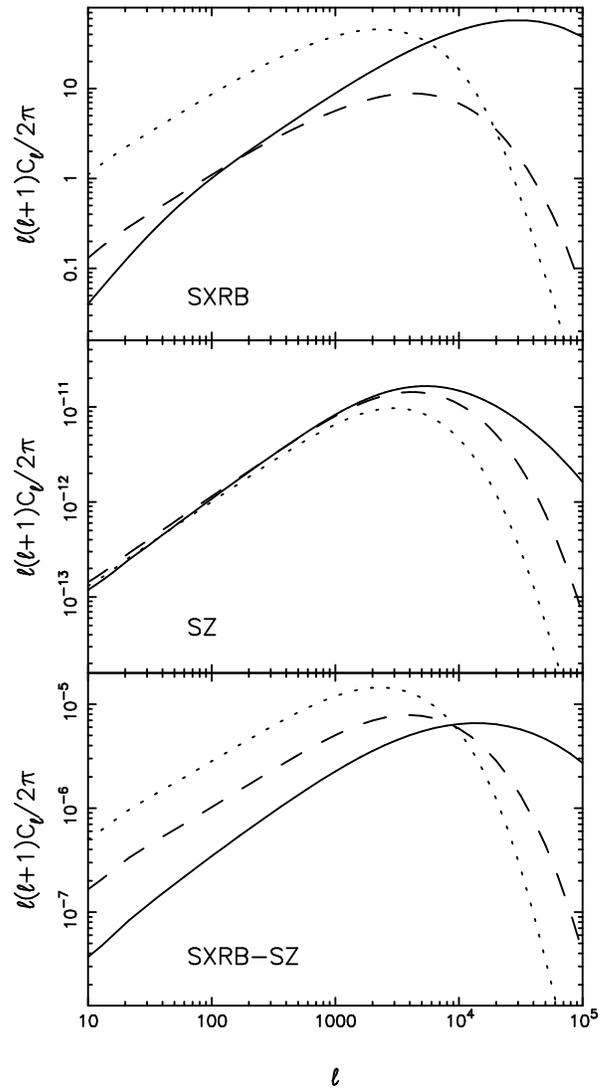}
\caption{Effect of non-gravitational processes on the power spectra
of SXRB-SXRB (top panel) , SZ-SZ (middle panel) and SXRB-SZ (lower panel)
correlations.
The results of self-similar model, preheating model and cooling model
are displayed by solid, dotted and dashed lines, respectively.}
\label{f3}
\end{figure}

In Figure~4, we demonstrate the expected cross power spectra 
before and after the clumping correction. The major uncertainty
in such an exercise is the clumping profile $C(r)$, which is poorly 
constrained by both current numerical simulations and X-ray observations. 
Guided by the result of N-body simulations (e.g. Ghigna et al. 2000;
Zentner \& Bullock 2003) 
that the spatial distribution of the clumps can be approximated 
by power laws, we take a toy model for $C(r)$ which has 
a power-law form of $C(r)=(\frac{r}{r_{\rm vir}})^\alpha$. We fix  
the power law index $\alpha$ through $C(r)=1.5$ at $r=0.2r_{\rm vir}$.
Of course, 
the oversimplification of this model can only give us a sense of 
how the cross power spectra might be affected by the gas clumping.  
We find that the difference is minor for our toy model. 
To obtain a substantial modification, which will be detectable, 
the clumping factor should be at the level of $\sim$ 10.

\begin{figure}

	\psfig{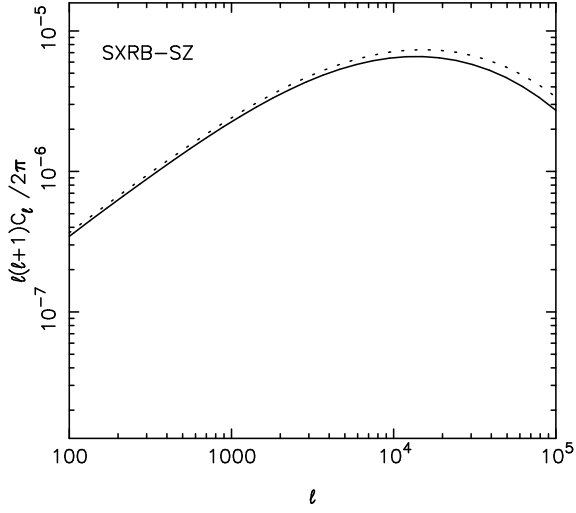}
\caption{Effect of the WHIM clumping on the cross power spectra.
Solid line corresponds to the WHIM distribution described by
the NFW profile without clumping structure [$C(r)=1$], and dotted line 
represents that with a clumping characterized by 
$C(r)=(r/r_{\rm vir})^{-0.25}$.}
\label{f4}
\end{figure}

In Figure~5, we plot the correlation coefficient between SZ and SXRB. 
As shown, the correlation coefficient has a value of $\sim$ 0.3, suggesting 
that SZ and SXRB are not well correlated. While this is partly due to
mismatches in redshifts where contributions arise, the
correlation coefficient can be as high as 0.9 when one considers 
contributions only at high redshifts. The correlation coefficient 
is mostly constant when $\ell$ ranges over three decades of 
magnitude between 100 an 10$^5$, suggesting
that the scale dependence between SZ and SXRB is not drastically different.

\begin{figure}
	\psfig{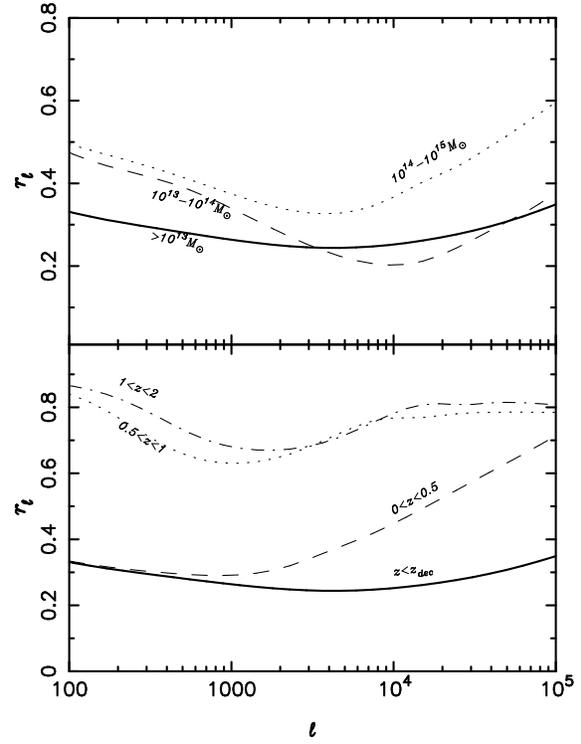}
\caption{The cross-correlation coefficient for the SZ effect and SXRB.
With no separation of contributions in either redshift or mass-space, 
the correlation is at the level of $\sim$ 0.3.}
\label{f5}
\end{figure}

We now demonstrate the possibility that 
the cross power spectrum of SXRB and SZ map can be detected with 
upcoming CMB experiment such as Planck and X-ray measurement such as  
DUTE mission. To describe errors related to an SZ map, we follow
the approach introduced by Cooray \& Hu (2000) and use the 
expected error on the SZ power spectrum based on multifrequency 
cleaning techniques.  
For the SXRB side, we make use of the X-ray catalog that is planned to
be produced with the DUET mission. This survey is expected to cover 
10,000 deg$^2$ down to a flux limit of $5 \times 10^{-14}$ 
ergs cm$^{-2}$ s$^{-1}$. The errors on the cross power spectrum
and the corresponding cross-correlation coefficient are 
calculated following the equations derived in Song et al. (2002).

We show in Figure~6 the expected cross power spectrum and  
correlation coefficient to be detected by Planck and DUTE missions. 
It turns out that with these future experiments one can determine 
the correlation down to $\ell\approx2000$ and  
to an accuracy of $\sim$ 0.04. However, 
the current CMB data, such as those from the WMAP satellite,
cannot be used to detect the cross-correlation 
between SZ effect and SXRB because 
the fluctuations associated with the SZ effect 
is dominated by the primordial anisotropies of CMB. 
For reliable studies, a separate SZ map is a must and an opportunity for this
is certainly available from the Planck mission.
While we have assumed a smooth distribution of gas within halos, 
any nonstandard physical effect, such as cooling and preheating,
will affect the cross-correlation. Though 
Planck and a similar X-ray catalog will present a first opportunity 
for a reliable detection of the cross-correlation between SZ 
and X-ray backgrounds, with targeted observations of small areas 
with better sensitivities, especially on the SZ side, we expect 
significant improvements and opportunities for understanding effects
such as non-gravitational heating.

\begin{figure}
	\psfig{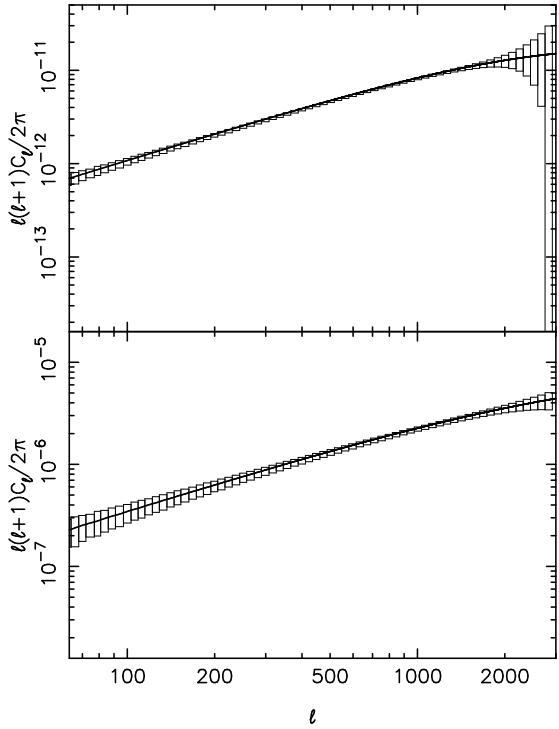}
\caption{Expected angular power spectrum of the SXRB-SZ correlation and 
corresponding correlation coefficient,
with no separation of contributions in either redshift or mass-space.
The error boxes are for the observation of the $25\%$ of the sky 
with the DUET mission and include the Plank noise. 
Bin width is chosen to be $\Delta \log \ell=0.025$.}
\label{f6}
\end{figure}

\section{Discussion and Conclusions}

Both X-ray emission and SZ effect arise from the WHIM 
gravitationally bound in massive clusters or large-scale structures.
However, the two phenomena
have very different response to the underlying gas distribution. 
The X-ray emission is more sensitive to the clumped gas structures 
(i.e. the central cores of  clusters), while the SZ effect probes 
a much wider region of gas distribution out to virial radii of the systems. 
This behavior is  reflected by the power spectra of their 
auto-correlation functions that are peaked at a larger $\ell$ 
for the SXRB and a smaller $\ell$ for the SZ map (see Fig.~1). Moreover, 
X-ray emission and SZ effect demonstrate different dependence
on the distances of clusters from us. As a consequence,  the major
contribution to the SXRB comes from  
nearby clusters. This compares to the thermal SZ sky at small
angular scales, which is dominated by high-redshift ($z>0.5$)
clusters. Therefore, the cross-correlation between 
the SXRB and SZ sky allows us to probe the distribution and 
evolution of the hot gas at intermediate angular scales and redshifts,
as are shown by Fig.~2. Even though the cross-correlation 
coefficient is relatively mild $\sim 0.3$ (Fig.~5),  
Planck can allow a reliable detection of that out to a multipole 
of $\sim 3000$. Further improvements one can hope will be achieved in the 
post-Planck era, and detection of such a correlation would allow us 
to further understand the gas/baryon distribution and the certain 
physical properties. 
Actually, cross-correlation between SXRB(or SZ) and 
galaxies have been extensively explored in literature. An corporation of 
these cross-correlation and auto-correlation analyses of the SXRB may 
constitute a powerful tool to expose the evolution and distribution of the 
missing baryons in the universe. 
Of course, our theoretical predictions have been made 
without correcting for various spurious correlations. 
For example,  most of the SXRB is actually generated by 
AGNs rather than diffuse sources like clusters and groups. 
Even if the contribution of AGNs can be nicely removed, the residuals  
may still be dominated by some nearby rich clusters (e.g. Diego et al. 2003). 
As for the SZ sky, potential sources of contaminations are the Milky Way,
radio point sources and even radio halos of clusters 
(Cooray et al. 1998; Bouchet \& Gispert 1999; Holder 2002;
Rubinn\~o-Mart\'in \& Sunyaev 2002; Zhou \& Wu 2003; etc.), provided
that primary CMB anisotropies are successfully subtracted.
Therefore, much work should be done to understand uncertainties 
in the detection of the SXRB-SZ cross correlation 
at small angular scales below $\sim10$ arcminutes in future experiments.

\begin{acknowledgements}
We gratefully acknowledge the valuable comments by an anonymous referee. 
This work was supported by the National Science Foundation of China, 
under Grant No. 10233040, and the Ministry of Science and Technology 
of China, under Grant No. NKBRSF G19990754.
\end{acknowledgements}

\end{document}